\documentclass{webofc}
\usepackage[varg]{txfonts}   
\newcommand{\beq}{\begin{eqnarray}}
\newcommand{\eeq}{\end{eqnarray}}
\woctitle{Husimi distribution for nucleon tomography}
\begin{document}
\title{Husimi distribution for nucleon tomography}
%
%

\author{Yoshitaka Hatta\inst{1}\fnsep\thanks{\email{hatta@yukawa.kyoto-u.ac.jp}} \and
        Yoshikazu Hagiwara\inst{2}
}

\institute{Yukawa Institute for Theoretical Physics, Kyoto University, Kyoto  606-8502, Japan
\and
     Department of Physics, Kyoto University, Kyoto 606-8502, Japan   
          }

\abstract{%
We define the QCD Husimi distribution as the phase space distribution of partons inside the nucleon. Compared to the more well-known Wigner distribution, the Husimi distribution is better behaved and positive. It thus allows for a probabilistic interpretation and can be used to define the `entropy' of the nucleon as a measure of complexity of the partonic structure. A possible connection to the Color Glass Condensate approach at small-$x$ is also discussed.   }
\maketitle
\section{Introduction}
\label{intro}

Over the past two decades, the study of the nucleon structure has literally expanded its dimensions. In addition to the ordinary parton distribution function (PDF) $f(x)$ which describes the one-dimensional (longitudinal) structure of the nucleon, the transverse-momentum-dependent distribution (TMD) $T(x,\vec{k}_\perp)$ and (the  Fourier transform of) the generalized parton distribution (GPD) $G(x,\vec{b}_\perp)$ have become common theoretical tools to reveal the three-dimensional imaging of the nucleon. Moreover, the five-dimensional Wigner distribution $W(x,\vec{k}_\perp,\vec{b}_\perp)$ \cite{wigner} has been conceived as the phase space distribution of quarks and gluons inside the nucleon  \cite{Belitsky:2003nz}. The Wigner distribution has the favorable property that, upon integration over one variable, it reduces to the known distribution in the other variable: $\int d^2\vec{b}_\perp W(x,\vec{k}_\perp,\vec{b}_\perp) = T(x,\vec{k}_\perp)$, $\int d^2\vec{k}_\perp W(x,\vec{k}_\perp,\vec{b}_\perp) = G(x,\vec{b}_\perp)$. On the other hand, it cannot be literally interpreted as a probability distribution in phase space because it is not positive definite and often behaves badly. This dilemma is actually well-known in nonrelativistic quantum mechanics. A less known fact  is that the Wigner distribution is not the unique phase space distribution in quantum mechanics, and there exists a positive definite distribution called the Husimi distribution \cite{husimi}.  In this contribution to the proceedings we discuss the potential use of the Husimi distribution in the context of the nucleon structure. For more details, see \cite{Hagiwara:2014iya}.

\section{Wigner and Husimi distributions in quantum mechanics}
\label{qm}

\begin{figure*}
\centering
\includegraphics[width=7cm,clip]{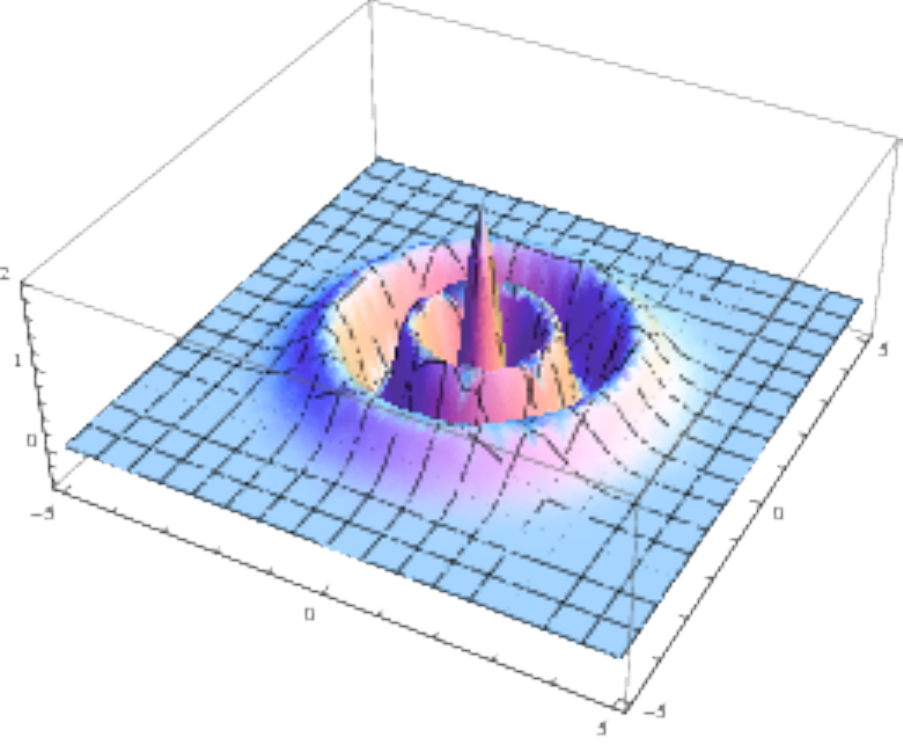}
 \includegraphics[width=6cm,clip]{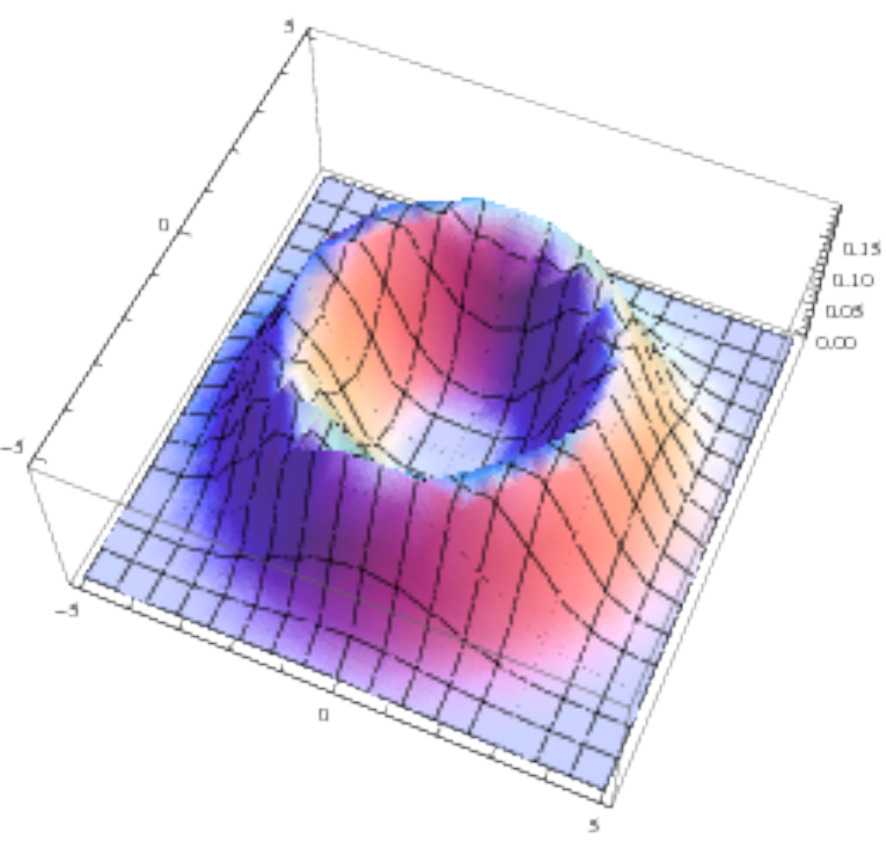}
 \caption{The Wigner distribution (left) and the Husimi distribution (right) in the $(q,p)$ plane of the 4th excited state of the one-dimensional harmonic oscillator ($m=\omega=\hbar=1$).}      
\label{fig2}       
\end{figure*}

Traditionally in quantum mechanics, the Wigner distribution \cite{wigner} has been most commonly used as a quantum analog of the phase space distribution $f(q,p)$ in classical mechanics. For a state $|\psi\rangle$, it is defined by
\beq
f_W(q,p)=\int dx e^{-ipx/\hbar}\langle \psi | q-x/2\rangle \langle q+x/2|\psi\rangle\,.
\label{def}
\eeq
 It immediately follows that
\beq
\int \frac{dq}{2\pi\hbar} f_W(q,p)=|\langle \psi |p\rangle |^2\,, \qquad
\int \frac{dp}{2\pi \hbar} f_W(q,p) = |\langle \psi |q\rangle |^2\,. \label{mom}
\eeq 
From these properties, it is tempting to interpret $f_W(q,p)$ as the probability distribution in phase space as in classical mechanics. However, this is not possible because $f_W$ is not positive definite.  We illustrate this point in the one-dimensional harmonic oscillator case for which the Wigner distribution can be analytically computed 
\beq
f_W(q,p)= 2(-1)^n e^{-\frac{2H}{\hbar\omega}} L_n\left(\frac{4H}{\hbar\omega}\right)\,, \label{wig}
\eeq
where $n$ is the excited level, $L_n$ is the Laguerre polynomial and $H=\frac{p^2}{2m}+\frac{m\omega^2 q^2}{2}$ is the classical Hamiltonian. Except for the ground state $n=0$, (\ref{wig}) is not positive definite, and as $n$ increases it oscillates more and more frequently. This is shown in Fig.~\ref{fig2}(left) for $n=4$. While the oscillations encode the quantum interference effects, the naive interpretation as a  phase space density drastically fails. The source of this difficulty is the uncertainty principle   
\beq
\Delta q \Delta p \gtrsim \frac{\hbar}{2}\,, \label{un}
\eeq
which tells that $q$ and $p$ cannot be determined simultaneously. Thus, in quantum systems, the very notion of `phase space distribution' is ill-defined from the outset. 

Nevertheless, it is physically reasonable to speak of a probability distribution in some averaged sense, as long as the averaging is done over a region in phase space larger than  (\ref{un}). A concrete realization of this idea is the Husimi distribution \cite{husimi} which is obtained by the Gaussian smearing of the Wigner distribution 
\beq
f_H(q,p) = \frac{1}{\pi \hbar}\int dq' dp'\, e^{-m\omega (q'-q)^2/\hbar -(p'-p)^2/m\omega\hbar} f_W(q',p')\,, \label{hu}
\eeq
 where the Gaussian widths in $q'$ and $p'$ are inversely related such that the product $\Delta q' \Delta p'=\hbar/2$ satisfies the minimal uncertainty. Remarkably, the Husimi distribution thus defined is positive (semi)definite. One can prove that 
 \beq
 f_H(q,p)= |\langle \psi |\lambda\rangle |^2\ge 0\,,
 \eeq
 where $|\lambda\rangle$ is the so-called coherent state which is the eigenstate of the annihilation operator $\hat{a}|\lambda\rangle = \lambda |\lambda\rangle$ and $\lambda=\frac{m\omega q+ip}{\sqrt{2\hbar m \omega}}$ is the eigenvalue. Together with the normalization condition $\int \frac{dqdp}{2\pi\hbar}f_H(q,p)=1$, the Husimi distribution can thus be legitimately interpreted as a probability distribution. Since its invention in 1940, the Husimi distribution has been used for numerous applications in various branches of modern physics \cite{lee}. In Fig.~\ref{fig2}(right), we plot the Husimi distribution of the harmonic oscillator
  \beq
 f_H(q,p)= \frac{1}{n!}e^{-\frac{H}{\hbar\omega}} \left(\frac{H}{\hbar\omega}\right)^n\,,
 \eeq
 for $n=4$. 
 The peak is localized around the classical trajectory $H\approx 4\hbar \omega$,  which makes perfect sense. It is very difficult to get this physical picture from the corresponding Wigner distribution. In the left figure of Fig.~\ref{fig2}, the first thing one notices is the peak in the center, and one might wonder if there is important physics going on there. But it is just an artifact of the particular definition of the Wigner distribution.

\section{Wigner and Husimi distributions in QCD}
\label{sec-1}

Let us now turn to QCD. The quark Wigner distribution is defined by \cite{Belitsky:2003nz,Lorce:2011kd}

\beq
W(x, \vec{b}_\perp, \vec{k}_\perp) \!= \!\!\int \frac{dz^- d^2z_\perp}{16\pi^3}\frac{d^2\Delta_\perp}{(2\pi)^2}e^{i(xP^+z^- -\vec{k}_\perp \cdot \vec{z}_\perp)}  \langle P+\!\Delta/2| \bar{\psi}(b -z/2)\gamma^+{\mathcal L}\psi(b +z/2)|P-\!\Delta/2\rangle \,, \label{wigQCD}
\eeq
 where $|P\rangle$ is the nucleon state and ${\mathcal L}$ is the U-shaped Wilson line which connects the two points $b\pm z/2$ via the light-cone infinity $z^-=\pm \infty$. The gluon Wigner distribution can be similarly defined. An important difference with respect to the nonrelativistic case (\ref{def}) is that the initial and final states are different due to the momentum recoil $\vec{\Delta}_\perp$ which inevitably occurs when probing a relativistic system. Integrating over $\vec{b}_\perp$, one find the TMD distribution
 \beq
\int d^2\vec{b}_\perp W(x, \vec{b}_\perp, \vec{k}_\perp) =\int \frac{dz^- d^2z_\perp}{16\pi^3}e^{i(xP^+z^- -\vec{k}_\perp \cdot \vec{z}_\perp)}  \langle P| \bar{\psi}(-z/2)\gamma^+{\mathcal L}\psi(z/2)|P\rangle =T(x,\vec{k}_\perp)\,.
\eeq 
The Wigner distribution has been computed in many models \cite{Belitsky:2003nz,Lorce:2011kd,Lorce:2011ni,Kanazawa:2014nha,Mukherjee:2014nya,Courtoy:2014bea,Mukherjee:2015aja,Liu:2015eqa}. In models based on constituent quarks without gluons, one typically obtains a smooth positive distribution. However, once gluons are included, it behaves very badly (see below). This motivates us to define the QCD Husimi distribution  
\beq
&& H(x, \vec{b}_\perp, \vec{k}_\perp)\equiv  \frac{1}{\pi^2}\int d^2b'_\perp d^2k'_\perp e^{-\frac{1}{\ell^2}(\vec{b}_\perp-\vec{b}'_\perp)^2 -\ell^2(\vec{k}_\perp-\vec{k}'_\perp)^2}W(x, \vec{b}'_\perp, \vec{k}'_\perp) \label{hu2} \\
&&=  \int \frac{dz^- d^2z_\perp}{16\pi^3} \frac{d^2\Delta_\perp}{(2\pi)^2}e^{i(xp^+z^- -\vec{k}_\perp \cdot \vec{z}_\perp)}e^{-\ell^2\frac{\Delta_\perp^2}{4}-\frac{z_\perp^2}{4\ell^2}} \langle P+\Delta/2| \bar{\psi}(b-z/2)\gamma^+ {\mathcal L}\psi(b+z/2)|P-\Delta/2\rangle  \,.  \nonumber
\eeq
As before, the widths of the two Gaussian factors are inversely related to each other.  The value of $\ell$ is in principle arbitrary, but there is a natural choice. To see this, note that the $\vec{b}_\perp$-integral of the Husimi distribution does {\it not}  reduce to the TMD distribution
\beq
\int  d^2\vec{b}_\perp H(x,\vec{b}_\perp, \vec{k}_\perp)= \int \frac{dz^- d^2z_\perp}{16\pi^3} e^{i(xp^+z^- -\vec{k}_\perp \cdot \vec{z}_\perp)} e^{-\frac{z_\perp^2}{4\ell^2}}  \langle P| \bar{\psi}(-z/2)\gamma^+ {\mathcal L}\psi(z/2)|P\rangle\,.
\label{gau}
\eeq
The extra Gaussian factor $e^{-z_\perp^2/4\ell^2}$ may seem worrisome, but actually it is not something completely unfamiliar. It can be viewed as an effective nonperturbative cutoff on the $z_\perp$-integral which mimics confinement. (See also a discussion at the end.) Indeed, if we approximate $\ell \gtrsim z_\perp\approx 0$ in the matrix element, we find
\beq
\int d^2\vec{b}_\perp H(x,\vec{b}_\perp,\vec{k}_\perp) \approx \frac{1}{\pi\langle k_\perp^2\rangle} e^{- \vec{k}_\perp^2/\langle k_\perp^2\rangle}f(x)\,, \label{mauro}
\eeq
 where $f(x)$ is the ordinary PDF and we identified $1/\ell^2 =\langle k_\perp^2\rangle$, the average transverse momentum squared. (\ref{mauro}) agrees with the factorized ansatz for the TMD which has been successfully employed in  phenomenology \cite{Anselmino:2013lza}. This suggests that $1/\ell$ should be of the order of the typical transverse momentum of partons.
 
The double moment of the Husimi distribution is the ordinary PDF. 
 \beq
\int \!d^2b_\perp d^2k_\perp H(x,\vec{b}_\perp,\vec{k}_\perp) = \int d^2b_\perp d^2k_\perp W(x, \vec{b}_\perp,\vec{k}_\perp)=f(x)\,.
\eeq 
Similarly, the canonical orbital angular momentum can be computed both from the Husimi and Wigner distributions \cite{Lorce:2011kd,Hatta:2011ku}
\beq
L_{can} = \int dx  d^2b_\perp d^2k_\perp (\vec{b}_\perp\times \vec{k}_\perp) H(x, \vec{b}_\perp, \vec{k}_\perp) =\int dx d^2b_\perp d^2k_\perp (\vec{b}_\perp\times \vec{k}_\perp) W(x, \vec{b}_\perp, \vec{k}_\perp)\,. \eeq

\section{One-loop example}

We now come to the issue of positivity. Actually, in the QCD case one cannot rigorously prove that the Husimi distribution as defined in (\ref{hu2}) is positive definite because it is not the forward matrix element due to the momentum recoil $\Delta_\perp$. However, the Gaussian factor $e^{-\ell^2 \Delta_\perp^2/4}$ suppresses large values of $\Delta_\perp$, and there is a good chance that (\ref{hu2}) can be treated as a positive function for practical purposes. Here we demonstrate this in a one-loop calculation. The one-loop Wigner distribution for a single quark (or an electron in QED, up to the Casimir factor) is \cite{Mukherjee:2014nya}
\beq
W(x,\vec{b}_\perp,\vec{k}_\perp)=\frac{\alpha_s C_F}{2\pi^2} \int \frac{d^2\Delta_\perp}{(2\pi)^2}
e^{-i\vec{\Delta}_\perp \cdot \vec{b}_\perp}
\frac{\vec{q}_+\cdot \vec{q}_- P_{qq}(x)+m^2(1-x)^3}{(q_+^2+m^2(1-x)^2)(q_-^2+m^2(1-x)^2)}\,, \label{ba}
\eeq
 where  $P_{qq}(x)=\frac{1+x^2}{1-x}$ is the splitting function and $\vec{q}_\pm \equiv \vec{k}_\perp\pm \frac{\vec{\Delta}_\perp}{2}(1-x)$.  The $\Delta_\perp$-integral can be done only numerically, but one already sees that (\ref{ba}) is a badly-behaved function.  Firstly, the $\Delta_\perp$-integral is divergent when $b_\perp=0$. One thus needs a cutoff and the result depends rather strongly on the cutoff. Second, the coefficient of $P_{qq}$ is negative when $k_\perp < \Delta_\perp$, and this makes $W$ itself negative for small $k_\perp$.  Furthermore, the factor $e^{-i\vec{\Delta}_\perp \cdot \vec{b}_\perp}$ oscillates rapidly at large $b_\perp$ and this leads to oscillations in $W$. 

\begin{figure*}[h]
\includegraphics[width=7.5cm]{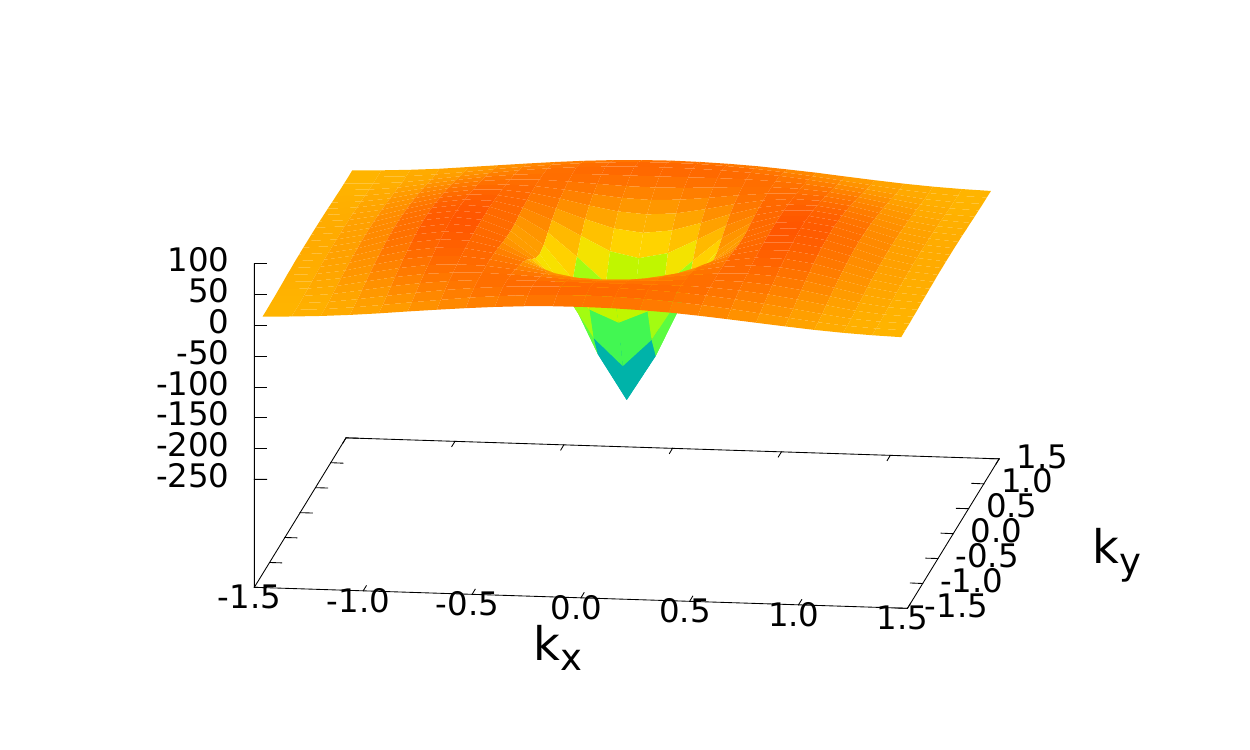}
 \includegraphics[width=7.5cm]{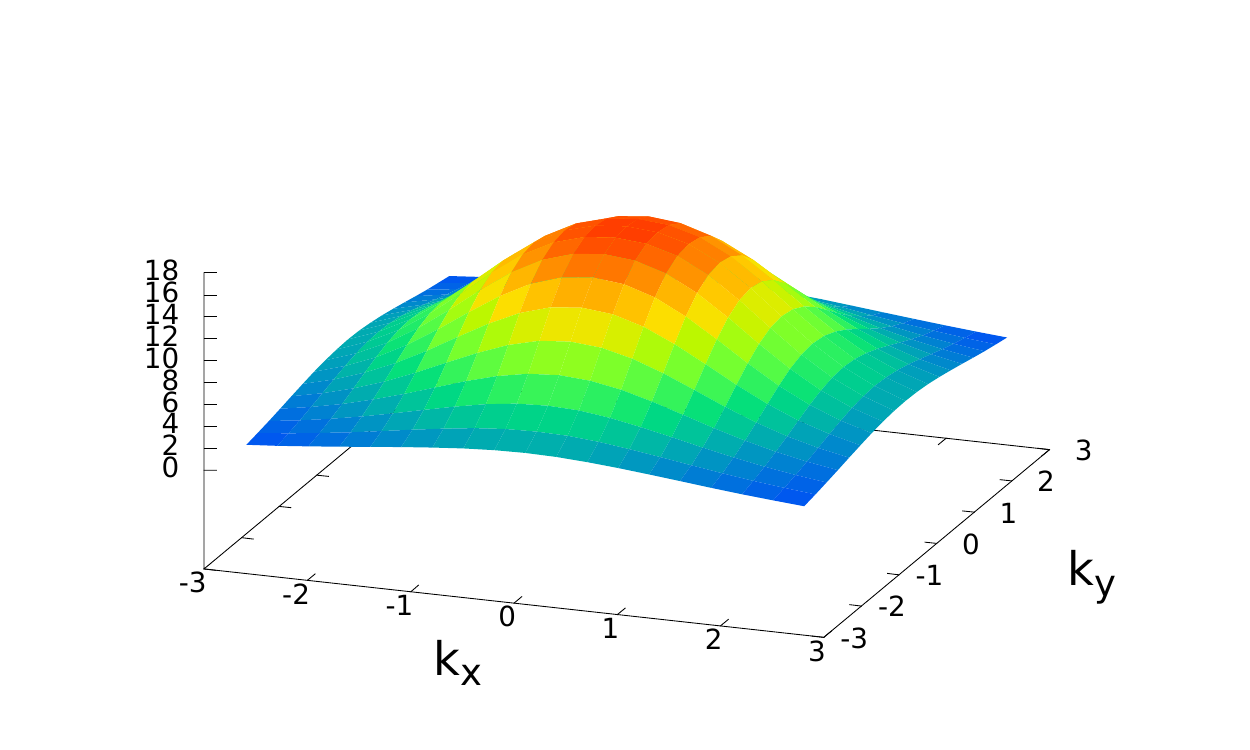}
 \caption{The one-loop Wigner distribution (left) and the Husimi distribution (right) for a single quark in the $\vec{k}_\perp$-space. We set $x=0.5$, $m^2=0.1\mbox{GeV}^2$ and $\ell = 1\mbox{GeV}^{-1}$. The units of $k_{x,y}$ are in GeV$^{-1}$. The Wigner distribution is computed with a cutoff $\Delta_\perp^{max}=5$GeV.  }
\label{fig-3}       
\end{figure*}

All these problems can be resolved by switching to the Husimi distribution (\ref{hu2}). The Gaussian factor provides an effective cutoff $\Delta_\perp <1/\ell$, so there is no problem of convergence. Moreover, the widths of the two Gaussian factors are deliberately related such that the smearing is done in a region larger than the region where the coefficient of $P_{qq}$ is negative. In Fig.~\ref{fig-3}, we plot the numerical result in the $\vec{k}_\perp$-space at fixed $\vec{b}_\perp=(0.5\mbox{GeV}^{-1},0)$ (see also \cite{Mukherjee:2014nya}). As expected, there is a sharp negative peak in the small-$k_\perp$ region in the Wigner distribution (left). This is hard to interpret intuitively because one is supposedly calculating the phase space density of the quark number $\sim \bar{\psi}\gamma^+\psi$, but there is no antiquark in this one-loop example. Thanks to the Gaussian smearing, even a function like this can be converted to a smooth positive Husimi distribution as shown in the right plot. We have surveyed different sets of parameters and found no evidence of negative regions as far as we could see. Thus, our construction is working, and the QCD Husimi distribution can be interpreted as the probability distribution of partons inside the nucleon.    

\section{Discussions}

We conclude with two speculative remarks which may be worth pursuing in future work. 

\subsection{Entropy}

In quantum mechanics,
since the Husimi distribution $f_H$ is positive definite, one can take the logarithm and define an entropy called the Husimi-Wehrl entropy \cite{wehrl}
\beq
S \equiv -\int \frac{dq dp}{2\pi \hbar} \, f_H \ln f_H\,.
\eeq
 This is the classical counterpart of the quantum (von Neumann) entropy $S_q=-\mbox{tr}\hat{\rho}\ln \hat{\rho}$ ($\hat{\rho}$ is the density matrix), but unlike the latter, it is nonzero even for a pure state $\hat{\rho}=|\psi\rangle \langle \psi|$. It thus serves as the quantitative measure of complexity (or `chaoticity') of pure quantum states. It is interesting to similarly define the `entropy of the nucleon' from the QCD Husimi distribution 
 \beq
S(x) \equiv -\int d^2b_\perp d^2k_\perp H  \ln H \,.
\eeq
This characterizes quantitatively how `complex' the nucleon wavefunction is due to the internal distribution of quarks and gluons at a given value of $x$. See  also  \cite{Kutak:2011rb,Peschanski:2012cw,Tsukiji:2015rra,Kovner:2015hga} for related discussions.

\subsection{Relation to Color Glass Condensate}

We have argued that the parameter $1/\ell$ should be of the order of the average transverse momentum $\langle k_\perp\rangle$ in the nucleon. When $x\sim {\mathcal O}(1)$, $\langle k_\perp\rangle$ is a nonperturbative scale, of the order of $\Lambda_{QCD}$. However, at small-$x$ and/or in a large nucleus, there arises a new perturbative scale, called the {\it saturation momentum} $Q_s(x)$, which sets the typical transverse momentum scale \cite{McLerran:1993ni}. An effective theory of the nucleon structure in this regime is the Color Glass Condensate (CGC). It is then natural to choose $1/\ell = Q_s(x)$.  One then notices that the Gaussian factor in (\ref{gau}) becomes formally identical to the so-called dipole $S$-matrix
\beq
S(z_\perp)= e^{-z_\perp^2Q_s^2(x)/4}\,,
\eeq
 which is frequently encountered in the calculational framework of the CGC.   This may not be a coincidence. The idea of the CGC is that at small-$x$ there are so many gluons that they can be effectively treated as a classical, coherent field. On the other hand, the Husimi distribution is the coherent state expectation value of the density matrix. In a sense, the Husimi distribution aims to maximize the classical aspects of a given quantum state. Thus the two approaches can be related. As the parton density grows at small-$x$, the classical description of the nucleon/nucleus becomes more appropriate. The matrix elements computed within the quasiclassical/CGC approach could be interpreted as the Husimi distribution.

\end{document}